\documentclass[epj]{svjour}
\usepackage{graphics,graphicx}
\begin{document}
\title{Magnetic Anisotropy of Deposited Transition Metal Clusters}
\subtitle{}
\author{S. Bornemann\inst{1},  J. Min\'ar\inst{1}, J. B. Staunton\inst{2}, 
        J. Honolka\inst{3}, A.Enders\inst{3}, K. Kern\inst{3} \and H. Ebert\inst{1}
}                     
%
%
\institute{Department Chemie und Biochemie, Ludwig-Maximilians-Universit\"at M\"unchen, 81377 M\"unchen, Germany 
      \and Department of Physics, University of Warwick, Coventry CV4 7AL, United Kingdom 
      \and Max Planck Institut for Solid State Research, 70569 Stuttgart, Germany}
\date{Received:  / Revised version: }
%
\abstract{
We present results of magnetic torque calculations using the fully relativistic
spin-polarized Korringa-Kohn-Rostoker approach applied to small Co and Fe
clusters deposited on the Pt(111) surface. From the magnetic torque one can
derive among others the magnetic anisotropy energy (MAE). It was found that
this approach is numerically much more stable and also computationally less
demanding than using the magnetic force theorem that allows to calculate the
MAE directly. Although structural relaxation effects were not included our
results correspond reasonably well to recent experimental data.
} 
\authorrunning{S. Bornemann, J. Min\'ar, J. B. Staunton \and H. Ebert}
\maketitle
\section{Introduction}
\label{intro}
In recent years, low-dimensional magnetic nanostructures on surfaces have been
the subject of intense experimental and theoretical research activities which are
driven by fundamental as well as practical interests. One of the central
questions for potential future applications is how the spin-orbit
coupling (SOC) induces specific magnetic properties such as orbital
magnetic moments and magnetic anisotropy. Another interesting issue is its influence on the
exchange coupling. Hereby, it is especially important to understand how these
properties evolve from single magnetic adatoms to submonolayer magnetic
particles. This very prominent role of SOC for such systems is also reflected 
in recent reviews \cite{BBB+05} as well as theoretical work \cite{NDP06}.

A system which has been intensively investigated experimentally
\cite{MBF+06,GRV+03,RCW+03,PRB04,GDM+02,GRC+05,B05,GDM+04} as well as
theoretically \cite{SBM+06,BMP+05,MBS+06,LSW03,XB06} in recent years is
Co/Pt(111) (used here as a short notation for Co clusters or nanostructures,
respectively, deposited on a Pt(111) substrate) as this is a
prototype to study the requirements on new high-density magnetic storage
materials. Earlier theoretical works studied only rather small Co clusters or
Co chains on Pt(111) \cite{SBM+06,BMP+05,MBS+06,LSW03}, whereas only recently
first qualitative results and trends based on a parameterised tight-binding
approach were published for deposited structures of up to 37 Co atoms
\cite{XB06}.
In our earlier works \cite{SBM+06,BMP+05,MBS+06} on the Co/Pt(111) system we
already described the evolution of the spin and orbital moments as well as the exchange
coupling for small clusters. What has been missing so far, for a complete
picture of the magnetic behaviour is the magnetic anisotropy energy
(MAE) for such systems. In principle, calculations of the MAE are
possible by applying the magnetic force theorem or by determining the total
energies as a function of the magnetisation direction. However, it turned
out that these procedures are rather delicate when dealing with deposited
clusters and one needs to take great care when taking band or total energy
differences. Therefore, we implemented a method to calculate the magnetic
torque directly from the electronic structure. Calculating the magnetic torque
for a sequence of directions of the magnetisation then gives access to the MAE.

In this present work we show first results for magnetic torque calculations 
of the already well investigated Co adatoms and dimers deposited on Pt(111) and compare
them with their Fe analogues. These investigations are complemented by calculations
for decorated clusters that allow to optimize the MAE. Our theoretical results
are then used to simulate magnetisation curves of an ensemble of Fe$_{\rm n}$ (n=1,2,3)
clusters on Pt(111), that are compared to recent experimental results.

\section{Computational Details}
\label{CD}
The calculations for the investigated cluster systems were done within the framework of spin
density functional theory using the local spin density approximation (LSDA)
with the parameterization given by Vosko, Wilk and Nusair for the exchange and
correlation potential \cite{VWN80}. The electronic structure is determined in a
fully relativistic way on the basis of the Dirac equation for spin-polarised
potentials which is solved using the Korringa-Kohn-Rostoker (KKR) multiple
scattering formalism \cite{Ebe00,TB-SPR-KKR}. This procedure consists of two
steps. First the Pt(111) host surface is calculated self-consistently with the
tight binding (TB) version of the KKR method using layers of empty sites to represent the
vacuum region. This step is then followed by treating the deposited clusters as a
perturbation to the clean surface with the Green's function for the new system
being obtained by solving the corresponding Dyson equation. This scheme is described in
more detail in earlier publications \cite{BMP+05,MBS+06}.

In all calculations the cluster atoms were assumed to occupy ideal lattice sites in the first
vacuum layer and no effects of structure relaxation were included. Therefore,
our results contain a systematic error and are strictly spoken not directly 
comparable with experimental data. Nevertheless, it could be
shown already in earlier work \cite{SBM+06,MBS+06,LSW03} on deposited clusters that this approach is
capable of reproducing systematic trends as well as achieving a reasonable
quantitative agreement with values found in experiment.
 
The $\alpha_{\hat{u}}$-component  of the torque vector $T_{\alpha_{\hat{u}}}^{(\hat{n})}$ 
on the magnetic moments oriented along the direction $\hat{n}$ was
calculated with an expression based on Lloyd's formula and perturbation 
theory \cite{SSB+06}

\begin{eqnarray}
\label{toreq}
T^{(\hat{n})}_{\alpha_{\hat u}} & = & -\frac{1}{\pi}\int dE \, f_{FD}(E) \\ 
         &   &  \mathrm{Im} \; i \sum_i tr(\underline{\tau}^{(\hat{n})}_{ii} \, 
	   [({\hat u} \cdot \underline{\hat J}) {{\underline{t}_i}^{(\hat{n})}}^{-1} - 
            {{\underline{t}_i}^{(\hat{n})}}^{-1} ({\hat u} \cdot \underline{\hat J})    ] ) \;.\nonumber 
\end{eqnarray}

Here $f_{FD}$ is the Fermi distribution function, $\hat{u}$ is the direction of
the torque vector and $\hat{J}$ is the total angular momentum operator. Finally,
the matrices ${\underline{t}_i}^{(\hat{n})}$ and
$\underline{\tau}^{(\hat{n})}_{ii}$ are the single site t-matrix and the site
diagonal scattering path operator, respectively, where $i$ (used as an index)
labels the atomic sites. As the calculations are done assuming the temperature
T=0K the Fermi distribution function is replaced by the theta function
$\Theta(E_F-E)$ with $E_F$ being the Fermi energy.
For all results shown below, the energy integral $\int dE$ occuring in Eq.~(\ref{toreq}) was
calculated on a rectangular complex energy mesh containing 64 points, while
using an angular momentum expansion up to $l_{\rm max}=2$ for all the occuring 
matrices.

Eq.~(\ref{toreq}) uses the analytic derivative of the energy with respect to
a rotation angle. We found that this approach is numerically much more robust
than taking the differences between band or total energies. The disadvantage, however,
is that the magnetic anisotropy energy, defined as the 
difference $E(\hat{n},\hat{n}_0)$ of the energy for two orientations of the magnetisation, $\hat{n}$
and $\hat{n}_0$, respectively, has to be determined by a corresponding path integral:

\begin{equation}
E(\hat{n},\hat{n}_0) = \int_{\hat{n}_0}^{\hat{n}} T^{(\hat{n})} d\hat{n} \;.
\label{Eint}
\end{equation}

Developing $E(\hat{n},\hat{n}_0)$ in spherical harmonics up to second order 
and taking into account the symmetry of the investigated cluster substrate
systems, one finds for example for a cluster having C$_{2\rm v}$-symmetry
with respect to its spatial structure, i.e. ignoring the orientation $\hat{n}$
of the magnetisation \cite{LUS+04}:

\begin{eqnarray}
E(\theta, \phi) & = &E_0 + K_{2,1}\cos{2\theta} \\ 
                &   & + K_{2,2}(1-\cos{2\theta})\cos{2\phi}+ 
                K_{2,3}\sin{2\theta}\sin{\phi}  \nonumber \;.
\end{eqnarray}

Using a corresponding expression for the torque it is straight forward to
deal with the integral occurring in Eq.~(\ref{Eint}). The evaluation of the
anisotropy constants $K_{n,m}$ occurring in this equation can then be determined
in a rather easy way by determining the torque for certain orentations $\hat{n}$
i.e. at angles $(\theta,\phi)$ of the magnetic moments (see below).

\section{Results and Discussion}
\label{RD}
\begin{figure}
\begin{center}
\resizebox{.8\columnwidth}{!}{\includegraphics[clip]{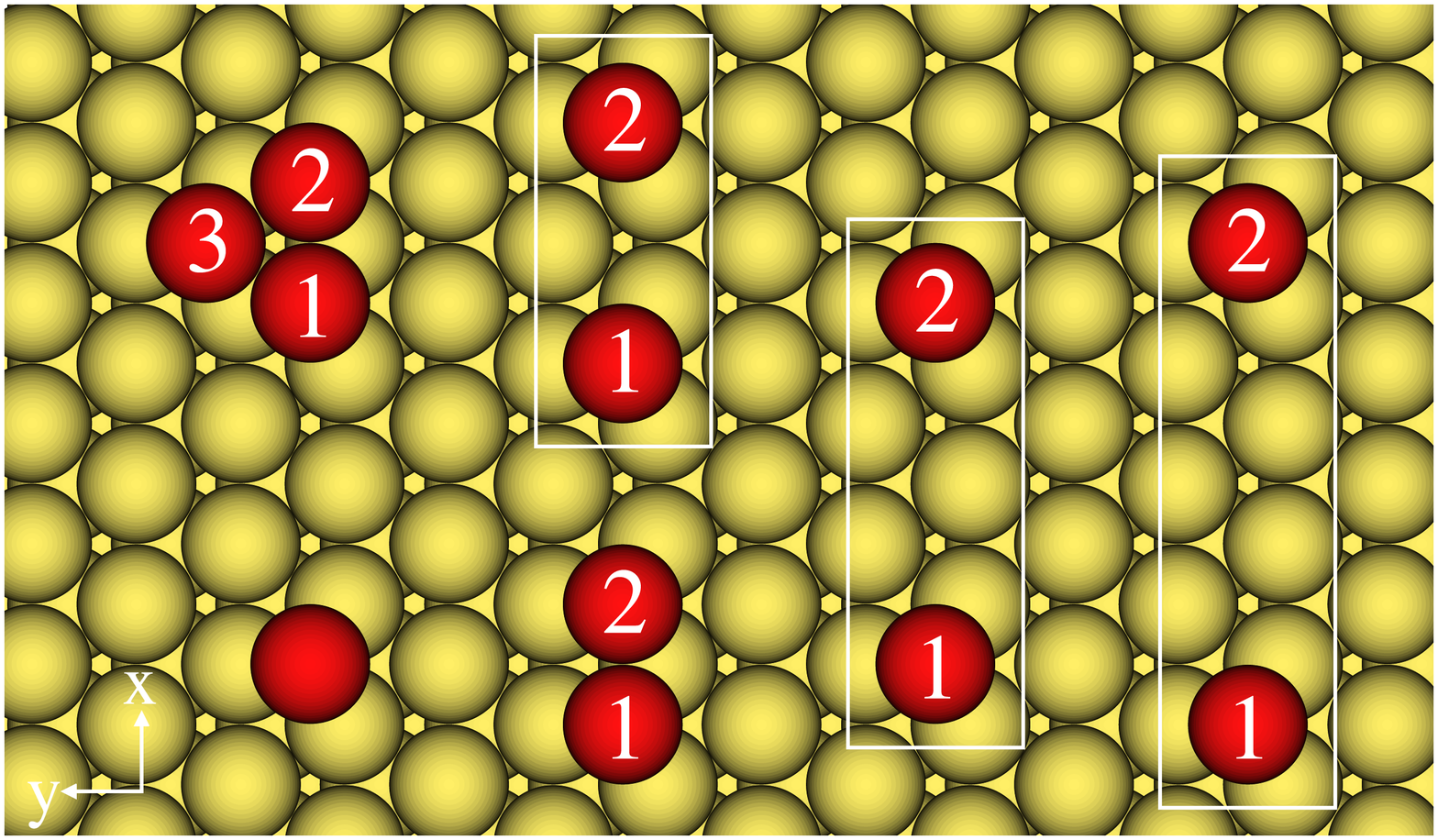}}
\end{center}
\caption{Structures of the investigated systems: The cluster atoms occupy
         ideal lattice sites of the underlying Pt(111) substrate.}
\label{fig:1} 
\end{figure}

The structure of the investigated Co and Fe monomers and dimers are shown together
with the underlying Pt(111) substrate in Fig. \ref{fig:1}. 
As the ad- or cluster atoms, respectively occupy regular lattice sites 
correspondingly to the substrate the resulting cluster/substrate system has
C$_{3\rm v}$- and C$_{2\rm v}$-symmetry, respectively.
Comparing the
resulting spin and orbital magnetic moments of Co and Fe in Tab. \ref{tab:1}
one notices that the spin magnetic moments for Fe are in general about 1.5
times larger than for Co.  For both transition metals the dimer formation has
only a minor impact on their spin magnetic moments when compared to the single
adatoms. The orbital moments, however, show a more interesting behaviour. Here
we find already a substatial quenching when going from single adatoms to the
corresponding pure i.e. unmixed dimers. This effect seems to be much more
pronounced in the case of Fe where the orbital magnetic moment of a Fe$_2$
dimer atom reduces to about one third of the monomer value compared to only
three quarters in the case of Co. The corresponding values for the mixed dimer
differ only slightly from those of Co$_2$ and Fe$_2$, respectively. The
different behaviour of the orbital moment for Co and Fe is also reflected in
their anisotropies (see below). A further increase in cluster size leads
usually (depending also on the cluster shape) to a  rapid and monotonous decay
of spin and orbital magnetic moments approaching rapidly the values of a
monolayer.

\begin{table}
\caption{Spin and orbital moments per atom of the investigated systems with magnetisation
         along the z-axis, i.e. perpendicular to the surface. For the mixed cluster the value
         refers to the component underlined.}
\label{tab:1}       
\begin{tabular}{lcccccc}
\hline\noalign{\smallskip}
& Co$_1$ & Fe$_1$ & Co$_2$ & Fe$_2$ & Fe\underline{Co} & \underline{Fe}Co  \\
\noalign{\smallskip}\hline\noalign{\smallskip}
$\mu_{\rm spin}$ ($\mu_{\rm B}$)& 2.27 & 3.49 & 2.15 & 3.33 & 2.12 & 3.38  \\
$\mu_{\rm orb}$ ($\mu_{\rm B}$) & 0.60 & 0.77 & 0.44 & 0.24 & 0.39 & 0.26  \\
\noalign{\smallskip}\hline
\end{tabular}
\end{table}

\begin{figure}
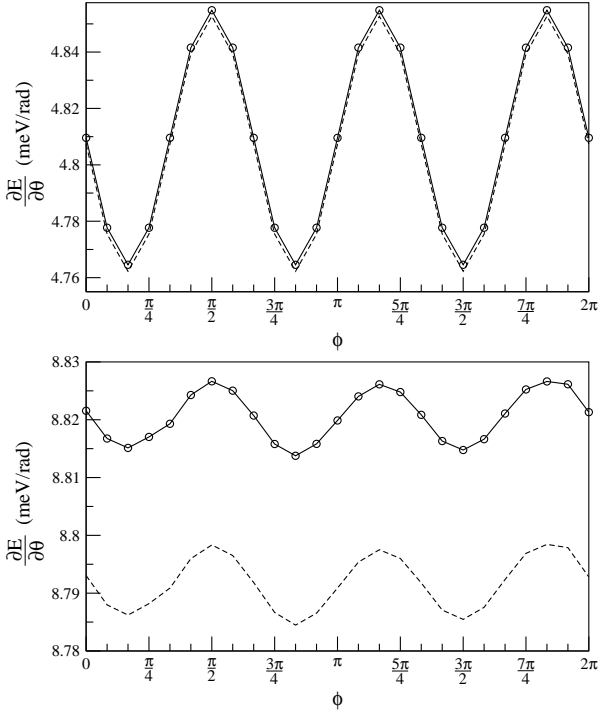

\begin{center}
\resizebox{0.9\columnwidth}{!}{\includegraphics[clip]{figs/Co_monomer.eps}}
\resizebox{0.9\columnwidth}{!}{\includegraphics[clip]{figs/Fe_monomer.eps}}
\end{center}
\caption{Torque component $T_{\theta}(\theta,\phi)$ for Co$_1$ (top) and Fe$_1$ (bottom) at 
         $\theta=\frac{\pi}{4}$ as a function of the polar angle $\phi$ (In fact 
         $-T_{\theta}(\theta,\phi)=\frac{\partial E}{\partial \theta}$ is shown -- see text).  
         The dashed line shows the same results including the substrate effect.}
\label{fig:monomer} 
\end{figure}

Fig. \ref{fig:monomer} shows the dependence of the $\theta$-component of the
magnetic torque $T_{\theta}(\theta,\phi) = - \frac{\partial E}{\partial
\theta}$ on the polar angle $\phi$ for Co and Fe monomers at
$\theta=\frac{\pi}{4}$.  
As $T_{\theta}(\theta,\phi)$ is found to be negative here as well as for the 
following we show always $\frac{\partial E}{\partial \theta}=-T_{\theta}(\theta,\phi)$.
The positive sign of $\frac{\partial E}{\partial \theta}$ for all angles $\phi$
implies that the torque forces the magnetisation to the z-axis. This means that
the system's easy axis points out of plane along $\hat{z}$.  One can see that the threefold
symmetry imposed by the underlying Pt substrate is directly reflected by the
small oscillations of $T_{\theta}$ with the polar angle $\phi$. 

This also demonstrates the high sensitivity of our implementation. The curves
show no numerical noise even for an energy resolution below 0.1 meV. As the
$\phi$ dependence of $T_{\theta}$ is so small when compared to its absolute
value the adatoms almost behave like perfect uniaxial magnets. In this case the
anisotropy constants $K$ can be extracted from the minima of
$T_{\theta}(\theta,\phi)$. For the single adatoms this gives then 4.76 meV for
Co and 8.79 meV for Fe. Fig.~\ref{fig:monomer} also shows the influence of the
induced anisotropy coming from the Pt substrate atoms. This induced MAE is
about -30 $\mu$eV for Fe and even smaller in the case of Co.  In fact it seems
to be negligible for very small Co clusters composed of only few atoms.
For larger two-dimensional Co clusters, however, we found that this induced
anisotropy becomes more important with increasing cluster size and can rise to
the same order of magnitude  as the contribution coming from the Co atoms
themselves.

\begin{figure}
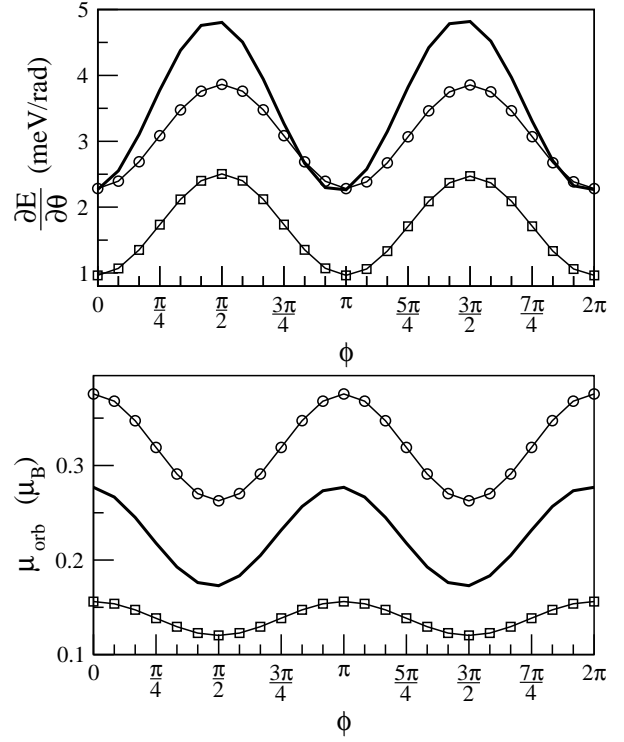

\begin{center}
\resizebox{0.9\columnwidth}{!}{\includegraphics[clip]{figs/dimers_torque.eps}}
\resizebox{0.9\columnwidth}{!}{\includegraphics[clip]{figs/dimers_orb.eps}}
\end{center}
\caption{Top panel: torque component $T_{\theta}(\theta,\phi)$ for Co$_2$ (circles), Fe$_2$ (squares) 
         and the FeCo dimer (thick line) for $\theta=\frac{\pi}{4}$. 
         Lower panel: $\phi$ dependence of the average orbital moments for 
         all three dimers at $\theta=\frac{\pi}{2}$.}
\label{fig:dimer} 
\end{figure}

The results of the torque calculations for the three different investigated 
dimers are shown in the top panel of
Fig.~\ref{fig:dimer}. In all cases $\frac{\partial E}{\partial \theta}$ is
again positive, however, the absolute values are significantly reduced when
compared to the monomers.  The $\phi$-dependence of the $\theta$-component of
the torque on the other hand is now much more pronounced due to the reduced
symmetry of the cluster/substrate system when compared to the monomer 
(see Fig.~\ref{fig:1}). From
Fig.~\ref{fig:dimer}  one can see that $E(\hat{n},\hat{z})$ is smallest if the
magnetic moments are oriented along the $\hat{x}$ direction, i. e. along the
cluster dimer axis (see Fig.~\ref{fig:1}).  The lower panel of
Fig.~\ref{fig:dimer} shows the corresponding $\phi$ dependence of the orbital
moments for $\theta=\frac{\pi}{2}$. One can see that the oscillations in the
orbital moments follow the oscillations of $T_{\theta}$ in an anticyclic
manner.  Here it should be pointed out that the largest values for the orbital
moments are obtained when the magnetisation points along the z-axis (see
Table~\ref{tab:1}). This as well as the behaviour seen in Fig.~\ref{fig:dimer}
is in qualitative agreement with the Bruno and van der
Laan anisotropy models \cite{Bru89,vdL98} that relate the MAE
$E(\hat{n},\hat{n}_0)$ to the corresponding anisotropy of the orbital moment.

\begin{figure}
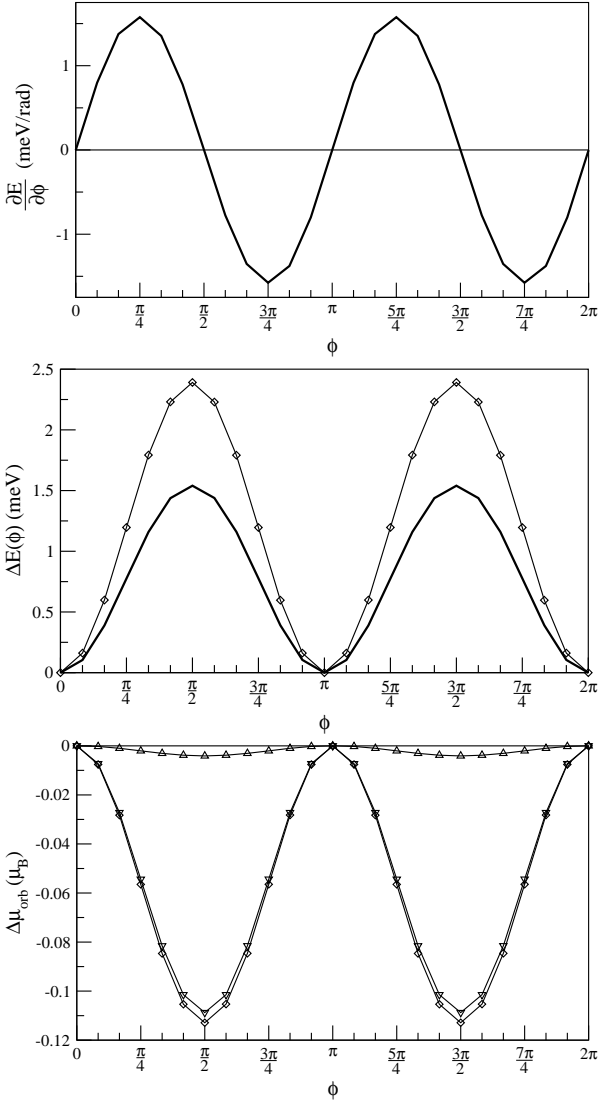

\begin{center}
\resizebox{0.9\columnwidth}{!}{\includegraphics[clip]{figs/Co2_dEdphi.eps}}
\resizebox{0.9\columnwidth}{!}{\includegraphics[clip]{figs/Co2_deltaE.eps}}
\resizebox{0.9\columnwidth}{!}{\includegraphics[clip]{figs/Co2_deltaMu.eps}}
\end{center}
\caption{Top panel: Torque component $T_{\phi}(\theta,\phi)$ of Co$_2$ 
         for $\theta=\frac{\pi}{2}$ as a fuction of $\phi$.
         Middle panel: MAE $\Delta E(\phi)$ for $\theta=\frac{\pi}{2}$ 
         refered to $\hat{n}_0\hat{=}(\frac{\pi}{2},\phi=0)$. The thick 
         line gives calculated results while the thin line marked by diamonds 
         gives results based on van der Laan's model. 
         Lower Panel: spin resolved anisotropy $\Delta\mu^{ms}(\phi)$ for 
         $\theta=\frac{\pi}{2}$ refered to $\hat{n}_0\hat{=}(\frac{\pi}{2},\phi=0)$
         for spin up and down (triangles up and down, resp.) and summed up (diamonds).}
\label{dEdphi} 
\end{figure}

For the van der Laan model one has 

\begin{eqnarray}
\label{vdL}
E(\hat{n},\hat{n}_0) & = & -\frac{\xi}{4\mu_B}[{\mu_{\rm orb}^{\uparrow}(\hat{n})} 
                           - {\mu_{\rm orb}^{\downarrow}(\hat{n})}] \\
                     &   & +\frac{\xi}{4\mu_B}[{\mu_{\rm orb}^{\uparrow}(\hat{n}_0)} 
                           - {\mu_{\rm orb}^{\downarrow}(\hat{n}_0)}] \nonumber  \;,
\end{eqnarray}

where $\xi$ is the appropriate spin-orbit coupling constant (here for the 3d
transition metal Fe or Co, respectively), and ${\mu_{\rm orb}^{m_s}(\hat{n})}$ is the
spin projected orbital magnetic moment for an orientation of the magnetisation
along $\hat{n}$.  If one considers a strong ferromagnet one can ignore the majority
spin contribution in Eq.~(\ref{vdL}) leading to the expectation that the orbital
magnetic moment takes its maximum for the magnetisation oriented along the easy
axis. The results shown in Figs.~\ref{fig:monomer}  and \ref{fig:dimer} are
obviously in full agreement with this. As Fig.~\ref{fig:dimer}  shows for the
three considered dimers, also the MAE deduced from the shown torque goes
parallel with the orbital magnetic moment (lower panel) when $\phi$ is varied
while $\theta$ is kept fixed. 

The relation of the MAE and the anisotropy of the orbital magnetic moment has
been studied in more detail by calculating the $\phi$-component of the torque
for $\theta$ fixed to $\frac{\pi}{2}$, i.e. for the magnetisation forced to lie
in the surface xy-plane. As one notes from the results given in
Fig.~\ref{dEdphi}, the variation of the torque is again quite pronounced and
reflects the C$_{2\rm v}$-symmetry of the dimer/substrate system.  Integrating
the torque component with respect to $\phi$ (see Eq.~(\ref{Eint})) one is led 
to the MAE $\Delta E((\frac{\pi}{2},\phi),(\frac{\pi}{2},\phi))$ shown in the middle
panel of Fig.~\ref{dEdphi}, that again shows that the energy is at its minimum
if the magnetisation lies along the dimer axis. The corresponding anisotropy
$\Delta \mu(\phi)$ is given in the lower panel of Fig.~\ref{dEdphi} in a
spin-polarized way. As one notes the majority contribution to $\Delta
\mu_{\rm orb}(\phi)=\mu_{\rm orb}(\frac{\pi}{2},\phi)-\mu_{\rm orb}(\frac{\pi}{2},0)$ 
is very small and negligible. This is a consequence of the nearly
complete filling of the majority spin d-band of the Co-atoms, i.e. their strong
ferromagnetic behaviour. As a consequence, one has not to distinguish between
Bruno's and van der Laan's models. Using the relation given in Eq.~(\ref{vdL})
together with the spin orbit coupling strength $\xi=85$ meV for Co
\cite{PEND01} one is lead to the estimate for the MAE $\Delta E(\phi)$ in the middle
panel represented by the curve marked by diamonds.
Obviously, the qualitative behaviour of the MAE is properly reproduced with a
reasonable quantitative agreement.  

Figs.~\ref{fig:monomer}  and
\ref{fig:dimer} show that going from a monomer to a dimer the MAE is
drastically reduced. This goes on with increasing cluster size, although there
is some influence of the cluster shape \cite{GRV+03}. As one notes, the decrease
of the MAE is much more pronounced for Fe than for Co. For the mixed FeCo
cluster, however, the MAE is quite high. This indicates that by a suitable
combination of atoms one may optimize the MAE while keeping the magnetisation
high. In fact, there are already a number of experimental studies done along
this line. Here, we show results for the impact on the properties of a Fe
trimer on Pt(111) due to a decoration with Pt atoms. 

\begin{figure}
\begin{center}
\resizebox{0.9\columnwidth}{!}{\includegraphics[clip]{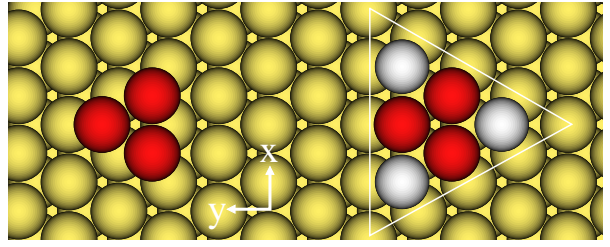}}
\end{center}
\caption{Structure of the studied Fe trimer on Pt(111) without (left) and
         with (right) decoration by three Pt atoms.}
\label{Tri_struc} 
\end{figure}
\begin{table}
\caption{Spin and orbital moments per atom of the Fe$_3$ and Fe$_3$Pt$_3$ 
         clusters on Pt(111) for the magnetisation 
         along the z-axis, i.e. perpendicular to the surface. The values
         refer to the component underlined. The last two columns give results
         for the Pt-substrate atoms having one or two Fe atoms as nearest
         neighbours, resp., in case of Fe$_3$Pt$_3$.}
\label{tab:2}       
\begin{tabular}{lcccccc}
\hline\noalign{\smallskip}
& Fe$_3$ &  \underline{Fe}$_3$Pt$_3$ & Fe$_3$\underline{Pt}$_3$ & Pt(1) & Pt(2) \\
\noalign{\smallskip}\hline\noalign{\smallskip}
$\mu_{\rm spin}$ ($\mu_{\rm B}$)& 3.20 & 3.16 & 0.21 & 0.09 & 0.11 \\
$\mu_{\rm orb}$ ($\mu_{\rm B}$) & 0.16 & 0.09 & 0.08 & 0.02 & 0.02 \\
\noalign{\smallskip}\hline
\end{tabular}
\end{table}

\begin{figure}
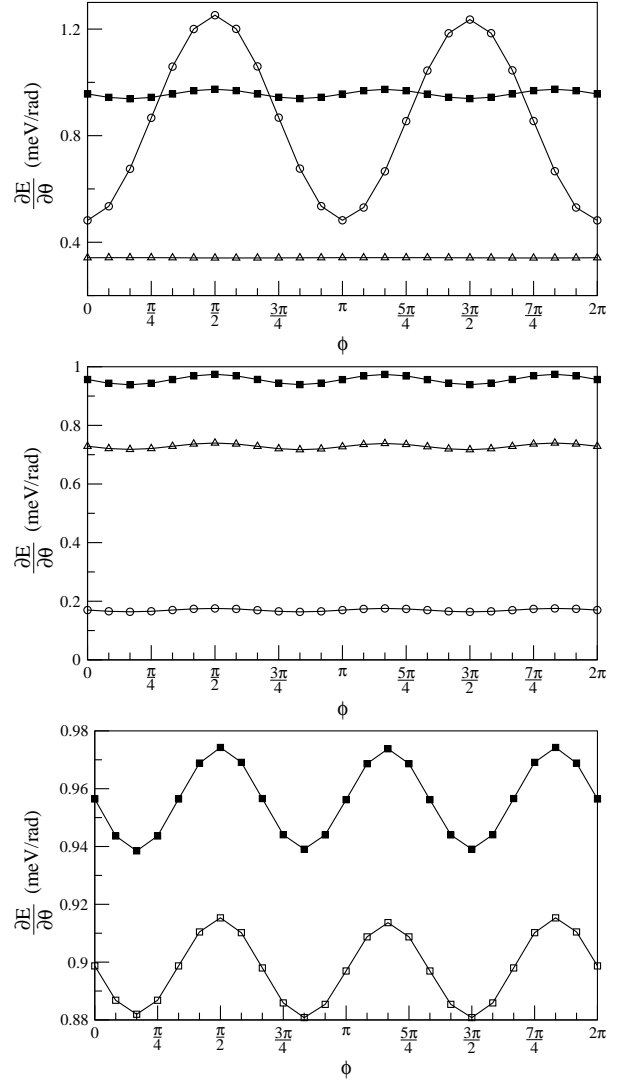

\begin{center}
\resizebox{0.9\columnwidth}{!}{\includegraphics[clip]{figs/Fe_trimer.eps}}
\resizebox{0.9\columnwidth}{!}{\includegraphics[clip]{figs/Pt3Fe_trimer_a.eps}}
\resizebox{0.9\columnwidth}{!}{\includegraphics[clip]{figs/Pt3Fe_trimer_b.eps}}
\end{center}
\caption{Top panel: torque component $T_{\theta}(\theta,\phi)$ 
         for Fe$_2$ (circles), Fe$_3$ (triangles)
         and Pt$_3$Fe$_3$ (squares) for $\theta=\frac{\pi}{4}$ as a function of the polar angle $\phi$.
         Middle Panel: torque component $T_{\theta}(\theta,\phi)$ per Fe atom for Pt$_3$Fe$_3$ at
         $\theta=\frac{\pi}{4}$ as a function of $\phi$: 
         total (full squares) including the substrate contribution, contribution of a
         Fe atom (triangle) and a cluster Pt atom (circles).
         Lower Panel: total torque component $T_{\theta}(\theta,\phi)$ per Fe atom for Pt$_3$Fe$_3$ at
         $\theta=\frac{\pi}{4}$ as a function of the polar angle $\phi$ with (full squares) and
         without (open squares) the substrate contribution.}
\label{fig:trimer} 
\end{figure}

Fig.~\ref{Tri_struc}  shows the corresponding structure of the investigated
cluster/substrate systems. The resulting spin and orbital magnetic moments are
given in Table~\ref{tab:2}. As one notes, the moments of the Fe atoms are again quite high
as for the mono- and dimer. For
the spin moment one finds only a minor influence due to the decorating Pt
atoms. These have also quite an appreciable induced magnetic moment. The
resulting torque $T_{\theta}(\theta,\phi)$ for $\theta=\frac{\pi}{4}$ as 
a function of the angle $\phi$ is shown in Fig.~\ref{fig:trimer} 
(top panel) for the Fe$_2$-, Fe$_3$- and Fe$_3$Pt$_3$ clusters. 
As is demonstrated once more the torque $T_{\theta}$ 
and with this also the MAE is strongly reduced going from the dimer to the
trimer. Adding the decoration, however, the high MAE of the dimer is
recovered. As the inner Fe$_3$ cluster of the Fe$_3$Pt$_3$ cluster is now surrounded by
neighbouring atoms the variation of the MAE with $\phi$ is strongly reduced when
compared with the dimer.

As one would expect, the MAE of the decorated trimer Fe$_3$Pt$_3$ is dominated
by its Fe contribution. This can be seen in the middle panel of
Fig.~\ref{fig:trimer} where apart from the total MAE per Fe-atom the averaged
contribution of a Fe-atom is shown.  Nevertheless, there is an appreciable contribution
coming from the decorating Pt atoms as well. This has to be ascribed on the
one-hand side to their appreciable induced magnetic moment (see Tab.~\ref{tab:2}) and on
the other side to their large spin-orbit coupling ($\xi_{\rm Pt}$=710 meV to be
compared with $\xi_{\rm Fe}$=65 meV). Apart from the contribution of the
decorating Pt atoms there is a noteworthy contribution from the neighbouring Pt-substrate
atoms as well. As can be seen in the lower panel of Fig.~\ref{fig:trimer} this
amounts to about 5\% of the total torque or MAE, respectively.

\begin{figure}
\begin{center}
\resizebox{0.9\columnwidth}{!}{\includegraphics[clip]{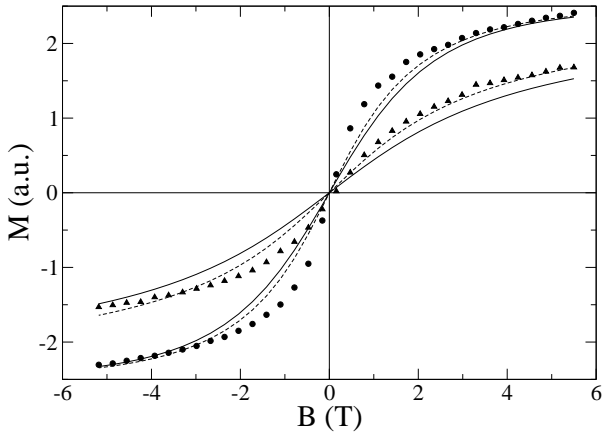}}
\end{center}
\caption{Experimental magnetisation curves $M(B)$ (dots) of an ensemble if 
         Fe$_n$ clusters on Pt(111) measured at T=6K for an orientation of the
         magnetic field $M(B)$ along the easy axis $\hat{z}(\theta=0^\circ)$
         and rotated by $\theta=65^\circ$ with respect to that. The full lines
         give corresponding theoretical results obtained on the basis of the 
         calculated  properties of Fe$_n$ clusters and the Langevin formula given
         in Eq.~(\ref{Lang}). The dashed line is obtained by including Fe$_4$
         clusters in the simulation.}
\label{magcurve} 
\end{figure}

The results for the undecorated Fe$_n$ clusters (n=1,2,3) shown in the 
lower panel of Fig.~\ref{fig:monomer} and the top panel of
Fig.~\ref{fig:trimer} allows us to make contact to corresponding experimental
investigations \cite{RLR+06}. The sample preparation technique used within these
investigations led to an ensemble of Fe$_n$ clusters that was dominated by clusters
of size n=1-3 with their statistical weight determined to be w$_n$=0.53, 0.54
and 0.33 \% (for technical details see Ref. \cite{GDM+02}). Fig.~\ref{magcurve}
shows the magnetisation curves $M(B)$ measured for this cluster ensemble at T=6K
for an orientation of the external magnetic field $\vec{B}$ along the easy axis
$(\hat{z})$ and at an angle $\theta\hat{=}65^{\circ}$ with respect to this axis.
With the magnetic moments and the anisotropy parameters for the Fe$_n$ clusters
available the magnetisation curves $M(B)$ can be simulated by means of the so-called
Langevin formula \cite{GRV+03}.
This way the thermal average of the z-component $m_{nz}(B,T)$ of the moment $m_n$
of an Fe$_n$ cluster can be expressed by: 

\begin{equation}
m_{nz}(B,T) = \frac{\int_{0}^{\pi}\sin\theta d\theta e^{-E(B,T,\theta)/kT} m_n \cos\theta}
                {\int_{0}^{\pi}\sin\theta d\theta e^{-E(B,T,\theta)/kT} }   \;.
\label{Lang}
\end{equation}

For the simulation the energy $E(B,T,\theta)$ was assumed to consist of its Zeeman
and anisotropy contributions

\begin{equation}
E(B,T,\theta) = Bm_n\cos\theta + K_1^n\sin^2\theta
\end{equation}

where for the later one an uniaxial behaviour has been assumed. The corresponding 
anisotropy constant $K_1^n$ has been deduced from the results for the MAE shown
above.
Adding the magnetisation curves for the Fe$_n$ clusters weigh\-ted by their 
statistical weight w$_n$ one is lead to the full lines shown in Fig.~\ref{magcurve}.
The additional dashed lines stem from a second simulation done including tetramers.
This indicates that a certain amount of larger Fe clusters are formed during
the preparation process as expected by statistics.

The good agreement of the simulated curves with experiment obviously demonstrates
that the complex magnetic properties of transition metal clusters can indeed be 
understood and described without adjustable parameters on the basis of the approach
used here. It also shows that inclusion of relaxation effects when calculating the
cluster properties should only slightly modify the numerical results. Nevertheless,
corresponding numerical works are in progress to determine the influence of lattice 
relaxations.

\section{Conclusion} 
\label{Conc}

We have presented results for the magnetic anisotropy of various Co/Pt(111) and
Fe/Pt(111) cluster/substrate systems calculated by the fully relativistic
KKR-approach. It was demonstrated that calculating the magnetic torque instead
of the magnetic anisotropy energy directly, is numerical very robust and
reliable. This is in particular reflected by the accuracy achieved for the
dependency of the torque on the polar angle $\phi$ as well as the results for the
substrate contribution.

As in previous work it was found that increasing the cluster size in general a
rapid decrease of the MAE occurs. However, it could be shown that by suitable
formation of compound clusters this drop can be compensated. In fact, the
combination of an element with large magnetic moments with one having large
spin-orbit coupling seems to be a promising approach.

Finally, it could be demonstrated that using the calculated cluster properties
the results of experimental magnetisation curves could be reproduced in a very
satisfying way confirming the adequateness of our approach as well as the
interpretation of the experimental findings.

\section*{Acknowledgement}
\label{Ack}
We acknowledge support by the Deutsche Forschungsgemeinschaft
within the Schwer\-punktprogramm 1153
{\em Cluster in Kontakt mit Oberfl\"achen: Elektronenstruktur
und Magnetismus}.

%
\bibliographystyle{epj}
\bibliography{ani}
%

\end{document}